\documentclass[aps,prl,showpacs,preprintnumbers,superscriptaddress,twocolumn,nofootinbib]{revtex4-1}
\usepackage{graphicx}  
\usepackage{dcolumn}   
\usepackage{bm}        
\usepackage{amssymb,amsmath}   
\usepackage{color}

\usepackage[Q=yes,pverb-linebreak=no]{examplep}

\newcommand{\mpt}{p\hspace{-0.465em}/ }
\newcommand{\vmpt}{\vec{p}\hspace{-0.465em}/ }
\newcommand{\aNLO}{{\sc\small MadGraph5}\Q{_}{\sc\small aMC@NLO}}

\begin{document}


\title{T-odd Asymmetry in W+jet Events at the LHC}

\author{Rikkert Frederix}
\email{rikkert.frederix@cern.ch}
\affiliation{PH Department, TH Unit, CERN, CH-1211 Geneva 23,
Switzerland}

\author{Kaoru Hagiwara}
\affiliation{KEK Theory Center and SOKENDAI, 1-1 Oho, Tsukuba, Ibaraki
305-0801, Japan}

\author{Toshifumi Yamada}
\email{toshifumi.y@gmail.com}
\affiliation{Department of Physics and Center for Mathematics and
Theoretical Physics, National Central University, Chungli, Taiwan 32001,
ROC}

\author{Hiroshi Yokoya}
\email{hyokoya@sci.u-toyama.ac.jp}
\affiliation{Department of Physics, University of Toyama, 3190 Gofuku,
Toyama 930-8555, Japan} 

\date{\today}

\begin{abstract}
W bosons produced at high transverse momentum in hadron collisions can
 have polarization along the direction perpendicular to the production
 plane, which is odd under na\"ive-T-reversal where both the
 three-momenta and angular momenta are reversed.
Perturbative QCD predicts non-zero polarization at the one-loop level,
 which can be measured as parity-odd components in the angular
 distribution of charged leptons from the decay of W bosons.
We perform a detector-level simulation with the
 generator \aNLO, and demonstrate that the asymmetry can be observed
 at the 8~TeV LHC with 20~fb$^{-1}$ of data.
If confirmed, it will be the first experimental measurement of the sign
 of the imaginary part of one-loop QCD amplitudes.
\end{abstract}

\preprint{CERN-PH-TH-2014-120, KEK-TH-1750, UT-HET-095}
\pacs{13.38.-b, 11.30.Er, 12.38.Bx, 14.70.Fm}

\maketitle


Na\"ive-T-reversal is a unitary transformation in which we
impose time-reversal on the initial and final states respectively, but
do not reverse the time direction from the initial to the final state. 
In CP-conserving theories like perturbative QCD, asymmetry under
na\"ive-T-reversal appears through the absorptive part of the scattering
amplitudes~\cite{rujula,kai0}, and hence offers a non-trivial test of
perturbative QCD at one and higher-loop levels.
Various tests have been proposed in the past, including asymmetries in
$\Upsilon $ decay into 3 jets~\cite{rujula}, $e^+e^-$ annihilation
production of 3 jets~\cite{ee}, neutrino (electro) production of 2
jets~\cite{kai0,Ahmed:1999ix}, Drell-Yan production of high-$q_T$
W-boson at hadron collisions~\cite{kai,Hagiwara:2006qe}, Z-boson decay
into 3 jets~\cite{Hagiwara:1990dx}, and top-quark radiative
decays~\cite{top}.
Although the predictions deserve much interests as probes of the
absorptive part of the loop-level QCD amplitudes, no experimental
confirmation has been made so far. 

In this paper, we consider the W+jets production at the LHC, 
\begin{align}
p \, p \ &\rightarrow \ W^+(\rightarrow \, l^+ \, \nu_l) \ + \ {\rm
 jets},
\label{wjet}
\end{align}
where $l$ denotes $e$ or $\mu$, in which 
T-odd effects that flip sign under na\"ive-T-reversal arise
in the parity-odd ($P$-odd) angular distributions of $l$ in the decay of
the W boson~\cite{kai,Hagiwara:2006qe}.
The following subprocesses contribute to the above process in the
leading order (LO): $ug\rightarrow W^+d$, $u\bar{d}\rightarrow
W^+g$, $\bar{d}g\rightarrow W^+\bar{u}$.
The differential cross section for the process can be expressed as,
\begin{align}
&\frac{{\rm d}\sigma}{{\rm d}q_T^2 {\rm d}\cos \hat{\theta} {\rm d}\cos
 \theta {\rm d}\phi}  = F_1 (1+\cos^2 \theta) + F_2 (1-3\cos^2
 \theta) \nonumber \\
 &\quad\quad
 + F_3 \sin 2 \theta \cos
 \phi + F_4 \sin^2 \theta \cos 2\phi + F_5 \cos 
 \theta \nonumber \\
 &\quad\quad 
 + F_6 \sin \theta \cos \phi + F_7 \sin \theta \sin \phi + F_8 \sin
 2\theta \sin \phi \nonumber \\
 &\quad\quad 
 + F_9 \sin^2\theta \sin 2\phi.
\label{xsection}
\end{align}
Here $(\theta, \, \phi)$ measures the direction of the $l^+$
 three-momentum in the W-boson rest-frame whose $y$-axis is taken
 perpendicular to the scattering plane\footnote{%
The z-axis can be chosen along the direction of the W momentum in the
 laboratory frame (the helicity frame), or along the direction which
 makes the same angle with the two beam momenta (the Collins-Soper frame
 \cite{cs}).
The results in this report do not depend on the choice of the z-axis.
},
$q_T$ denotes the transverse momentum of the W boson, and $\hat{\theta}$
 denotes the scattering angle of the W boson in the W+jet center-of-mass
 frame. 
The structure functions $F_{1-9}$, which are functions of $q_T$ and
 $\cos\hat{\theta}$, are described by the polarization density 
 matrix for W+jet production.
The $F_1$ term governs the overall normalization, while the other eight
 terms affect the lepton angular distributions.
The LO analytical expressions for $F_{1-6}$ at ${\mathcal O}(\alpha_s)$
 are found in Ref.~\cite{chaichian}, and the 
 next-to-leading order (NLO) corrections have been analyzed in 
 Ref.~\cite{Mirkes:1990vn}. 
$F_{7-9}$ terms represent the $P$-odd and T-odd components of
 the lepton angular distribution, because under parity transformation
 or na\"ive-T-reversal, $\phi$ flips sign while $\hat{\theta}$ and
 $\theta$ remain unchanged.
The LO contribution to these terms comes from the absorptive part of the
 one-loop amplitudes at ${\mathcal O}(\alpha_s^2)$, whose analytical
 expressions are found in Ref.~\cite{kai}\footnote{%
The contributions from CP-violating terms in the Standard Model are
negligibly small.}.
Experimentally, some of the $P$-even azimuthal angular distributions have
been measured in W+jet events at the Tevatron~\cite{Acosta:2005dn} in
good agreement with the NLO QCD prediction~\cite{Mirkes:1990vn}. 
At the LHC, only the polar angular distributions have been
measured~\cite{Chatrchyan:2011ig,ATLAS:2012au}, which confirm the
helicity fraction of W bosons predicted in QCD~\cite{Bern:2011ie}.
In the rest of this work, we focus on the $F_7$ term, owing to the fact
that this has the largest size of asymmetry among the three
terms~\cite{kai,Hagiwara:2006qe}.

Although a simulation study at the parton-level indicates that the
 Tevatron has enough potential to observe the T-odd
 terms~\cite{Hagiwara:2006qe}, no experimental measurement has been
 reported so far.
One of the reasons for the difficulty of the measurement might be that
 loop-level effects, such as T-odd asymmetries of the amplitudes,
 were not available in the LO event generators which are commonly used to
 simulate detector responses by experimentalists. 
In this Letter, we study how the T-odd effects are included in
the multi-purpose NLO event generator \aNLO~\cite{amc}\footnote{%
We have confirmed by the stand-alone matrix-element calculation that the 
T-odd terms completely agree with the analytic
expressions in Ref.~\cite{kai} at arbitrary phase-space points.}, 
 which has been made public very recently. Furthermore, we demonstrate
 how the effects of QCD initial-state/final-state radiation (ISR/FSR)
 and those of finite 
 detector resolution affect the measurements.
In order to study systematics of higher order QCD corrections, 
 we prepare two types of event samples; one 
 is generated by \aNLO~\cite{mg5, amc} where the W+jet events are 
 calculated at the NLO+PS (parton shower) level, and
 the other is generated by a hand-made event generator which we call
 {\sc\small LOMC} where all the $F_{1-9}$ structure functions are
 implemented at the LO with the help of {\sc\small BASES/SPRING}
 code~\cite{Kawabata:1995th}. 
We stress that, although the \aNLO\ code generates events with NLO
 accuracy, the T-odd observables constructed from these events
 are accurate at LO because these observables receive
 contributions only at the one-loop level and beyond.

We remind the reader that all contributions to NLO calculations are
completely automated in the \aNLO\ code: the virtual corrections are 
computed in the {\sc\small MadLoop} module~\cite{Hirschi:2011pa},
which is based on the OPP integrand-reduction
method~\cite{Ossola:2006us} (as implemented in {\sc\small
  CutTools}~\cite{Ossola:2007ax}) and the OpenLoops
technique~\cite{Cascioli:2011va}; the factorization of the infra-red
singularities is achieved by adopting the FKS
method~\cite{Frixione:1995ms}, as implemented in the {\sc\small
  MadFKS} module~\cite{Frederix:2009yq}; and the consistent matching
to parton showers is obtained by using the MC@NLO
technique~\cite{Frixione:2002ik}.

For the \aNLO\ simulation, we generate the $pp\to\mu^+\nu_{\mu}j$
process\footnote{%
We do not take into account decays into $\mu^-$ and $e^\pm$, 
 but these can be used to collect more data or to check the results
 independently.} 
 at the NLO.
{\tt CTEQ6M} parton distribution functions (PDFs)~\cite{cteq6m} are used,
 and the factorization and renormalization scales are set to $\mu_F=\mu_R=q_T$.
Phase-space cuts are applied at the generation-level, which are
$q_T>25$~GeV, ${p_j}_T>25 $~GeV, ${p_\mu}_T>22$~GeV in the regions of
$|\eta_\mu|<2.5$, and ${p_\nu}_T>10$~GeV, where ${p_i}_T$ and $\eta_i$
are the transverse momentum and pseudo-rapidity of a particle $i$,
respectively. 
Parton showering and hadronization are simulated with
 {\sc\small Herwig6}~\cite{herwig6}, and detector simulation is performed with
 {\sc\small PGS4}~\cite{pgs4}. Jets are reconstructed using the anti-$k_T$ jet
 clustering~\cite{Cacciari:2008gp} with $\Delta R=0.4$.

We generate net about 100M of events with \aNLO\ as a difference between
 positive weight events and negative weight events.
The scale variation can be estimated at no extra computational
cost~\cite{Frederix:2011ss}. 
For the {\sc\small LOMC}, we perform the simulation in a similar
setup to that for \aNLO, but with {\tt CTEQ6L} PDFs and LO matching with
 parton showers. 
For each of the three choices of the scales, $\mu=q_T$, $q_T/2$ and $2q_T$,
 we generate 100M of only positive weight events.

For the generated events, we apply the following selection cuts.
Denoting the missing transverse momentum by $\vmpt_T$ and defining
the transverse mass as $M_T \equiv \sqrt{2({p_{l}}_T\,\mpt_T -
{\vec{p_l}}_T \cdot \vmpt_T)}$, we require
(a) one $\mu^+$ with $p_{T} > 25$ GeV and $|\eta|<2.4$;
(b) $\mpt_T > 25$ GeV;
(c) $q_T \equiv \vert {\vec{p_\mu}}_T + \vmpt_T \vert >
       30$ GeV;
(d) $M_T > 60$ GeV;
(e) the leading jet satisfies $p_{T} > 30$ GeV and $|\eta| < 4.4$.
After these selection cuts, the cross section is about 200~pb at the NLO.
We note that these cuts are similar to those applied in the earlier W
boson observation at the LHC~\cite{Chatrchyan:2011ig,ATLAS:2012au},
where a good signal-to-background ratio has been achieved.

To observe the $F_7$ contribution, we have to measure
$\sin\theta\sin\phi$ and $\cos\hat\theta$, event by event, because $F_7$
is an odd function of $\cos\hat\theta$.
We define the charged-lepton momentum component perpendicular to the
scattering plane as
\begin{align}
 p_l^\perp=\frac{\vec{p}_{p_1}\times\vec{q}_T\cdot\vec{p}_l}
 {|\vec{p}_{p_1}\times\vec{q}_T|},
\end{align}
 where $\vec{p}_{p_1}$, $\vec{q}_T$ and $\vec{p}_l$ are the right-moving
 proton momentum, the W transverse momentum and the lepton momentum,
 respectively, all in the laboratory frame. 
In terms of $p_l^\perp$, $\sin\theta\sin\phi$ of eq.~(\ref{xsection})
can be observed as, 
\begin{align}
 (\sin\theta\sin\phi)_{\rm obs.} = p_l^\perp/(m_W/2) \equiv x_l^\perp,
\end{align}
 in the narrow width limit of the W boson.
On the other hand, the measurement of $\cos\hat\theta$ is affected
by the two-fold ambiguity in determining the neutrino longitudinal
momentum, or the W-boson rest frame.
Instead, we use the pseudo-rapidity difference between the charged
lepton and the leading hard jet, $\Delta\eta \equiv \eta_{\mu}-\eta_j$,
which has strong correlation with
$\cos\hat{\theta}$~\cite{Hagiwara:2006qe}. 

\begin{figure}[tb]
  \begin{center}
   \includegraphics[width=0.4\textwidth]{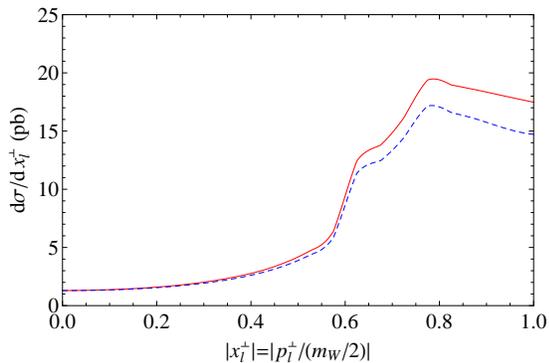}
  \end{center}
 \caption{
 $x_l^\perp=p_l^\perp/(M_W/2)$ distributions for the W+jet events after
 the selection cuts~(a-e) and a cut of $\Delta\eta> 1.0$,
 at the 8~TeV LHC, in the leading-order calculation at the parton-level.
 Predictions for $x_l^\perp>0$ and $x_l^\perp<0$ regions
 are separately plotted in the red solid and
 blue dashed lines, respectively. } 
 \label{sintsinp}
\end{figure}
\begin{figure*}[th]
 \begin{center}
  \includegraphics[width=\textwidth]{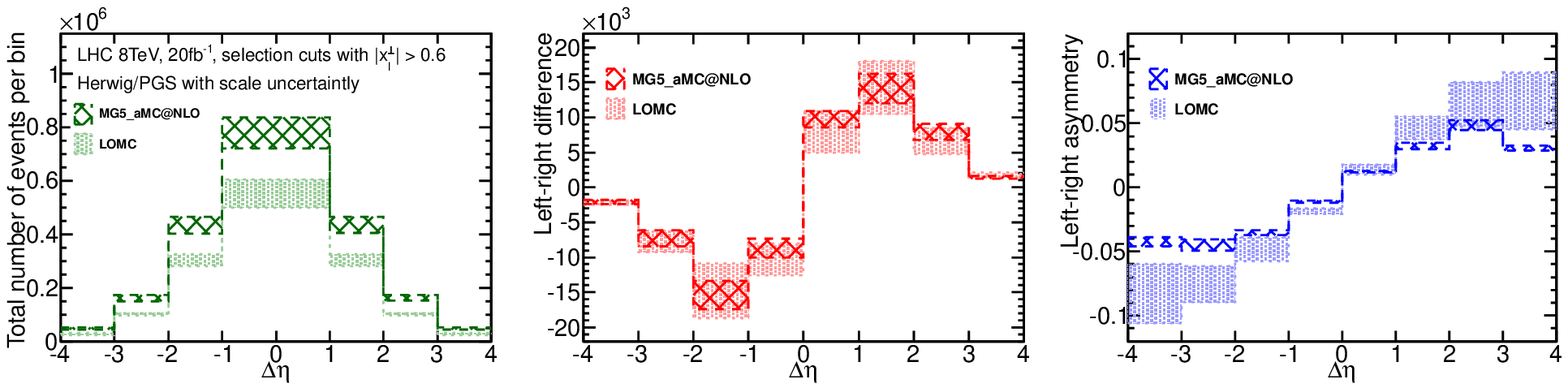}
  \caption{$\Delta\eta$ distributions for the cross section (left),
  left-right difference of the cross section (middle), and the
  left-right asymmetry (right) at the 8~TeV LHC after the selection cuts
  and a cut of $|x_l^\perp|>0.6$.
  Results by the {\sc\small MadGraph5}\_{\sc\small aMC@NLO} and
  {\sc\small LOMC} simulations are shown in dark-colored and
  light-colored histograms, respectively, with scale uncertainties.
  }\label{fig:dist}
 \end{center}
\end{figure*}

The determination of $x_l^\perp$ is affected by the uncertainty in the
$\vmpt$ measurement, because the scattering plane is determined by the W
transverse momentum, which is the vector sum of the lepton and missing
transverse momenta. 
To reduce the impact of this uncertainty, we select events with large
$|x_l^\perp|$ and simply focus on difference in the numbers of events
for $x_l^\perp>0$ and $x_l^\perp<0$, which we call the
\textit{left-right asymmetry}. 
To pin down an appropriate selection cut on $|x_l^\perp|$,
 we investigate the distribution of $x_l^\perp$.
In Fig.~\ref{sintsinp}, we show the $x_l^\perp$ distribution
 after the selection cuts~(a-e) and a cut of $\Delta\eta>
 1.0$ at the parton level, where an outgoing parton is identified with
 a hard jet. 
By selecting events with large $|x_l^\perp|$, we can reduce the smearing
 of the asymmetric distribution without loss of statistics.

Now that we have established the size of the asymmetry at the parton
 level, we present our main results in Fig.~\ref{fig:dist}.
In this figure, we show our simulated cross sections at the
 detector-level after the selection cuts~(a-e) and a cut of $|x_l^\perp|
 > 0.6$. 
The left, middle and right panels show
the $\Delta\eta$ distributions for the cross section, 
the left-right difference of the cross sections defined as
$\sigma(x_l^\perp>0)-\sigma(x_l^\perp<0)$,
and the left-right asymmetry of the cross sections defined as
\begin{align}\label{eqasym}
 A\equiv\frac{\sigma(x_l^\perp>0)-\sigma(x_l^\perp<0)}
{\sigma(x_l^\perp>0)+\sigma(x_l^\perp<0)},
\end{align}
respectively.
Results obtained by \aNLO\ and {\sc\small LOMC} simulations are shown
in the dark-colored large-hatched histograms and light-colored
small-hatched histograms, respectively.
Histograms are normalized to the expected number of events per bin at
the 8~TeV LHC with 20~fb$^{-1}$ of data after the selection cuts~(a-e) and
a cut of $|x_l^\perp|>0.6$ are applied.
The vertical widths of the histograms indicate the scale uncertainty in
the simulation.

As seen in the left panel, there is a difference between the predicted
cross sections for \aNLO\ and {\sc\small LOMC}.
This comes from nothing but the NLO correction to the total
cross section, which is included in the \aNLO\ but not in the
{\sc\small LOMC}. 
For our central scale choice, the $K$-factor is found to be around 1.5
for smaller $|\Delta\eta|$, but above 2 for larger $|\Delta\eta|$.
In the middle panel, the left-right difference of the cross sections 
is found consistent with the behavior of the $F_7$ terms.
The results by the two simulations are very similar, which is
consistent with our na\"ive expectation. This is because both the generators
contain the leading $\mathcal O(\alpha_s^2)$ terms for the $P$-odd
contributions. 
In principle, differences can be induced due to the use of different
set of PDFs and the different treatment in the parton-shower
simulation at the NLO and LO. 
However, our results suggest that these effects are negligibly small.
In the right panel, we find that an order of 5-10\% left-right asymmetry
is predicted and that the asymmetry is robust even after the inclusion
of QCD ISR/FSR and the detector smearing.
We point out that a smaller left-right asymmetry is predicted by \aNLO\
than {\sc\small LOMC}, due to the large enhancement of the total cross
section which enters in the denominator of the asymmetry. 

The scale uncertainties in our simulations deserve extra attention.
In the {\sc\small LOMC} simulation, there are two sources of scale
uncertainty; one is the choice of the scales at the parton-level
calculation, namely, the scales in strong coupling constant and in the
PDFs, and the other is the choice of the initial scale in the parton
showering. 
Variation of the choice of the former scales affects the cross section
and the lepton distribution at the parton-level.
Since the $P$-even ($P$-odd) part of the cross section is
${\mathcal O}(\alpha_s)$ [${\mathcal O}(\alpha^2_s)$], we expect an
overall scale dependence of ${\mathcal O}(10\%)$ [${\mathcal O}(20\%)$]
by the scale variation of $\alpha_s$. 
Variation of the scale in the parton showering affects the number and
distribution of the ISR/FSR jets.
In our event analysis, it affects the probability that 
the leading jet is misidentified by an ISR jet, which results in
the error of $\Delta\eta$.
If the scale is taken higher, more jets are produced via parton
showering, and the misidentification probability increases.
This causes a significant scale dependence in the cross sections for large
$|\Delta\eta|$, because ISR jets tend to appear at large
$|\eta_j|$.
The total scale uncertainty in the cross section is
not very large because the increase in the number of W+jet events due to
ISR jets is partially canceled by the smaller $\alpha_s$ at the higher
scale. 
For the left-right difference of the cross section, the shower-scale
variation does not cause significant shift in any $\Delta\eta$ regions, 
because the sum of the left-right difference over the entire
$\Delta\eta$ range is zero.
Its scale dependence is only governed by the overall
$\alpha_s^2$ factor.
Overall, the scale uncertainty in the left-right asymmetry is estimated
to be about 20\% (30\%) in the small (large) $|\Delta\eta|$ regions.

In the \aNLO\ simulation, there is a 
cancellation in the dependence on the parton shower starting scale and
the Monte Carlo subtraction terms~\cite{Frixione:2002ik}~leading to a
negligible uncertainty coming from this scale for the observables
studied here.
Therefore, the total scale uncertainty for the left-right asymmetry,
which is about 10\% in any region of $|\Delta\eta|$, is significantly
reduced from that in the {\sc\small LOMC} results.

The difference in magnitude of the left-right asymmetry between the
two simulations can be understood by the $K$-factor for the total
cross section in the \aNLO\ result, entering only in the denominator
of eq.(\ref{eqasym}). 
The {\sc\small LOMC} predictions do not have this apparent mismatch,
since both numerator and denominator are computed at LO accuracy. 
Since the difference between the \aNLO\ and {\sc\small LOMC}
simulations is larger than the accuracy of either one of these codes, we
regard this difference as an additional source of uncertainty for this
observable. 
To improve the situation, the NLO corrections also to the
numerator of eq.(\ref{eqasym}) are needed, however, they are currently
not known. 

Before closing, we present several comments.
We estimate the expected statistical error as $\delta
A=\sqrt{(1-A^2)/N_{\rm evt}}$, and find that with 20~fb$^{-1}$ of data,
$\delta A$ is about (1.1, 1.5, 2.5, 4.5)$\times10^{-3}$ for
$|\Delta\eta|=$([0,1], [1,2], [2,3], [3,4]) bins, respectively.
Therefore, the data collected at the LHC should be enough to
measure the asymmetry.
When a cut of $|x_l^\perp|>0.8$ is applied, the asymmetry is enlarged
by 10-20\%, while the statistical error also grows by about 30\%. 
We comment on background events from the $W^+\to\tau^+\nu$ decay followed by
the $\tau^+$ decay into $\mu^+$.
We find that such events do not exceed 2\% of the $W^+\to\mu^+\nu$
events in each bin of $\Delta\eta$ after selection cuts~(a-e) and a cut on $|x_l^\perp|>0.6$ are applied. 
Hence non-zero value of the left-right asymmetry is still observable in the presence of the $W^+\to\tau^+\nu$ background.

To summarize, we have examined the possibility of
observing T-odd asymmetry in W+jet events at the LHC.
The asymmetry arises from the absorptive part of the
scattering amplitudes in perturbative QCD, and manifests itself as a
difference in the parity-odd distributions in the lepton decay angle.
We have demonstrated by a simple detector-level analysis that the
difference due to the T-odd term remains detectable after
the inclusion of ISR/FSR radiation and detector resolution.
The prediction by the next-to-leading order event generator \aNLO\ 
contains relatively small scale uncertainties due to the matching to the
parton shower at the NLO accuracy.
On the other hand, the size of the asymmetry may be under-predicted, 
because the as-yet unavailable NLO corrections to the T-odd
cross section could be as large as those to the T-even cross
section. 

The work of HY was supported in part by Grant-in-Aid for Scientific
Research, No.\ 24340046.

\end{document}